**Elucidating the role of hyperfine interactions on organic magnetoresistance using deuterated aluminium tris(8-hydroxyquinoline)**

N.J. Rolfe[1], M. Heeney[2], P.B. Wyatt[3], A.J. Drew[1], T. Kreouzis[1], W.P. Gillin[1]

[1]Dept of Physics, Centre for Materials Research, Queen Mary University of London, Mile End Road, London, E1 4NS, United Kingdom

[2]Department of Chemistry, Imperial College London, Exhibition Road, London, SW7 2AZ, United Kingdom

[3]School of Biological and Chemical Science, Centre for Materials Research, Queen Mary University of London, Mile End Road, London, E1 4NS, United Kingdom

Measurements of the effect of a magnetic field on the light output and current through an organic light emitting diode made with deuterated aluminium tris(8-hydroxyquinoline) have shown that hyperfine coupling with protons is not the cause of the intrinsic organic magnetoresistance. We suggest that interactions with unpaired electrons in the device may be responsible.

In recent years there has been an increasing interest in the study of magnetic field effects on charge transport and recombination in organic light emitting diodes (OLEDs). Initial work by Kalinowski *et al.* [1] showed that, for aluminium *tris*(8-hydroxyquinoline) (Alq$_3$) based devices, it was possible to increase the current through a device by ~2.5% whilst improving the device efficiency by ~3%. These effects were observed at modest magnetic fields of less than 500 mT with the majority of the effect having occurred at fields of less than 50mT. These results have now been replicated in a number of organic molecular and polymeric systems [2-4] and are thought to be an intrinsic property universal across organic semiconductors. The effect of a magnetic field on the current has been dubbed organic magnetoresistance (OMR). However, there is still considerable discussion as to the precise mechanism behind the effect, with models based on either excitonic [1,5-9] or bipolaron [10] effects.

Despite the discussion about the details of the mechanism responsible for OMR, all the potential models rely on some degree of spin dynamics, with a common suggestion that spin-carrying radicals (polarons, excitons) are affected by hydrogen hyperfine fields [1,5,10-13]. These assumptions have been largely based on the observation that OMR occurs at very low magnetic fields, which are commensurate with those expected for hyperfine interactions. Despite this assumption, there has been little work to actually prove that hyperfine interactions are responsible. Nguyen *et al.* [13] tried to deduce the role of hyperfine coupling in OMR by studying device structures based on C$_{60}$, which contains no hydrogen. In their work they produced C$_{60}$ devices with a range of electrodes. Devices consisting of ITO/C$_{60}$/Ca/Al or Au/C$_{60}$/Ca/Al showed no OMR, whereas C$_{60}$ devices, using the highly doped conducting polymer poly(3,4-ethylenedioxythiophene)

poly(styrenesulfonate) (PEDOT) as the anode, demonstrated a definite positive OMR. The authors suggested that, despite control devices which consisted of PEDOT only showing a negative OMR, the OMR found for the PEDOT/$C_{60}$ device was not due to the $C_{60}$. It was therefore concluded that hyperfine interactions were the likely cause of OMR. We note that the absence of OMR in a given voltage range is not proof that the phenomenon can never occur in a particular material; there is still some debate as to the mechanism behind OMR and hence the conditions necessary to observe it. For example, the excitonic models suggest that exciton formation is an essential prerequisite for OMR and hence device structures with poor electron and/or hole injection will not show the effect. Indeed we have investigated OMR in the "hole-transport" material *N,N'*-diphenyl-*N,N'*-bis(3-methylphenyl)-(1,1'-biphenyl)-4,4'-diamine (TPD). For ITO/TPD/Au structures, which would be expected to be predominantly hole transport devices [14], no OMR could be observed below 7 V, whereas by replacing the gold with a better electron-injecting contact, such as aluminium, the onset of OMR could be seen at 1.4 V and weak electroluminescence could be observed at 3.5 V even though the power efficiency of the device was only ~$10^{-7}$%.

In light of this we have attempted to elucidate the role of hyperfine coupling due to hydrogen atoms on OMR by producing devices using deuterated aluminium *tris*(8-hydroxyquinoline) (Alq$_3$-*d$_{18}$*). If hyperfine interactions with hydrogen are the dominant cause of spin flipping, then by replacing the spin ½ hydrogen atoms in the active layer with spin 1 deuterium we should significantly perturb the observed OMR response. Any differences between the devices in terms of their efficiency or magnetoresistance would help show at which stage hyperfine interactions were occurring.

Deuterated 8-hydroxyquinoline was synthesised by a variation on the procedure of Tong *et al* [16]. A Teflon coated high pressure bomb containing a Teflon stirrer bead was charged with 8-hydroxyquinoline (1 g), $D_2O$ (13 ml), acetone-$d_6$ (2 ml) and a Pd/C catalyst (10% Pd, 0.5 g). The bomb was then heated in an oil bath at a temperature of 200°C with stirring at an estimated pressure of ~30-40 bar for 48 hours. The deuterated 8-hydroxyquinoline product was isolated and then recrystallized from hexane and characterised by mass spectrometry and [1]H and [13]C NMR spectroscopy. This material was then used to fabricate Alq$_3$-$d_{18}$ using a reaction with aluminium chloride in a methanol/water mix using an ammonia buffer. The resultant Alq$_3$-$d_{18}$ was purified by vacuum sublimation, at ~$10^{-7}$ mbar at a temperature of ~230°C, and characterised by mass spectrometry to ensure there was no proton exchange during the synthesis. The mass spectrometry of this material showed that it was 97% deuterated.

Devices were grown on an ITO coated glass substrate with a sheet resistivity of ~13 $\Omega/\square$ and consisted of a 50 nm TPD layer as the hole transport layer and 90 nm of Alq$_3$ or Alq$_3$-$d_{18}$ as the electron transporting/emissive layer with a cathode of 1nm LiF and 100 nm of aluminium. The ITO substrate was patterned using photolithography and cleaned by sequential ultrasonication in detergent solution, water, acetone and chloroform. Following this, the ITO was treated in an oxygen plasma for 5 min at 30 W and 2.5 mbar pressure using a Diener electronic femto plasma system. The plasma-treated substrate was immediately transferred to the deposition chamber for device fabrication. The deposition of the organic layers and metal electrodes were performed using a Kurt J. Lesker SPECTROS evaporation system with a base pressure during evaporation of ~$10^{-7}$ mbar. The rate of deposition of organic materials was about 0.2 nm/s while that of the aluminium was varied from ~0.1 to 0.5 nm/s. A calibrated oscillating quartz crystal monitor was used to determine the rate and thickness of the deposited layer. The whole device fabrication was performed without breaking vacuum.

Immediately after growth, the devices were placed in a light-tight sample holder with a calibrated silicon photodetector (Newport 818-SL) placed on the top surface of the device. The sample holder was placed between the poles of an electromagnet with the magnetic field perpendicular to the direction of current flow in the device. The photodetector was tested under various illumination levels to make sure there was no field dependence on its output. Measurements were taken with the device operated in constant voltage mode. Before and after each field measurement, a measurement at null field was taken. These two readings were averaged and used to remove any effects due to device drifting, thus allowing us to determine the effect of the magnetic field. Voltage sourcing and current measurements were performed using a Keithley 236 source-measure unit with current measurements being averaged over 32 readings. The optical power output was measured using a Newport 1830 optical power meter.

Figure 1 shows the current-voltage characteristics of an $Alq_3$ and an $Alq_3$-$d_{18}$ device, with nominal layer thicknesses the same. Photoluminescence spectra of the $Alq_3$-$d_{18}$ and $Alq_3$ materials are also shown in the inset to Figure 1; these are identical to the electroluminescence results of Tong et al [16] who also found no change in peak position with deuteration. The two devices clearly show identical performance, within experimental variations, in both their current voltage and luminance characteristics with both devices having maximum external power efficiencies of ~0.33%. This is in contrast to Tong *et al.* [16] who found that their deuterated devices showed significant improvements. However, their devices did not include a hole injection layer, which results in very high drive voltages and low current densities compared to our devices. They also only purified their material under a $10^{-3}$ Torr vacuum and then recrystallized it in ethanol. This may mean that it was less pure than the material we have used.

Figure 2 shows the OMR (percentage change in current) as a function of magnetic field for the two devices at a number of drive voltages. The correlation between the $Alq_3$ and $Alq_3$-$d_{18}$ devices is quite remarkable; the OMR for the two devices both have an almost identical shape and magnitude. For the deuteration levels we have achieved in this work (97%) it can be calculated that there are on average ~0.5 residual hydrogen atoms on each molecule. This strongly suggests that hydrogen hyperfine fields in $Alq_3$ are not responsible for OMR.

In addition to affecting the current through the device, the magnetic field also changes the device efficiency, as we have observed in our previous work on OMR [7-9,14,17]. The cause of OMR in OLEDs is still under debate. However, under the excitonic models it is suggested that the increase in efficiency is caused by a change in the relative population of singlets and triplets, and that this interaction is frequently cited as being of a hyperfine scale and hence due to interactions with proton spins [1,5,10-13].

Figure 3 shows the percentage change in efficiency for the $Alq_3$ and $Alq_3$-$d_{18}$ devices over the same range of voltages as in Figure 2. Again it can be seen that the curves are remarkably similar, but that the maximum change in efficiency is up to ~20% bigger in the $Alq_3$ device than for the $Alq_3$-$d_{18}$. This is within the variation that we find for nominally identical devices. Whether one considers that the change in efficiency with magnetic field is caused by mixing between the singlet and triplet states at either the pair state or excitonic level, it is clear from these results that neither the change in efficiency or the OMR are likely to be due to hyperfine interactions with protons.

It could be argued that we still have hydrogen in the TPD hole-transport layer in our device structure and we should therefore consider the role this may play. However, the proton-electron hyperfine coupling that is cited in the excitonic models is dominated by the Fermi contact term and/or the dipolar term, which are both very short ranged [18]. We would therefore expect hyperfine interactions from the TPD to only play a role in, at most, the first monolayer or so of the $Alq_3$-$d_{18}$. Given that the change in OMR is unaffected by deuteration and the fact that the emission, and hence the change in efficiency, in $Alq_3$ is known to come from a layer up to ~20nm thick [8] it is unlikely that the TPD interface is having any effect. We therefore conclude that interactions with proton spins are not the primary mechanism for the spin dynamics responsible for the observed change in OMR or efficiency; it is therefore necessary to consider other possible mechanisms.

Hyperfine interactions with other nuclei, such has [13]C, or impurities are unlikely to be responsible, since they are dilute and would be much weaker than those due to protons. Hence, if removing the protons shows no effect, other nuclear spin interactions can also be ruled out. Spin-orbit (SO) coupling could also a possibility but this is typically weak for the light elements found in organic semiconductors [19]. In our previous work, we found that changing the atomic mass of the central ion in the quinolate system from aluminium to indium has virtually no effect on the change in efficiency with applied magnetic field [17]. Since there is a well-known heavy atom effect to SO coupling [20], if it were an important ingredient behind OMR, more than quadrupling the mass of what is already the heaviest ion in the molecule should result in a significant increase in the interaction strength. It is therefore safe to conclude that SO coupling is not responsible for the mixing between triplet and singlet states under the influence of a magnetic field.

In light of this we suggest that interactions with paramagnetic species may be the dominant factor. Although these could be trapped charges, the relatively low concentration of these in addition to the short range of any coupling make this unlikely. Of far more relevance would be the interaction with free electrons (polarons). The magnetic moments of electrons are almost three orders of magnitude greater than for protons and one would therefore expect them to have a strong influence on spin decoherence, since the isotropic hyperfine coupling constant varies with the product of the two magnetic moments [21]; therefore the interaction would be expected to be a factor of ~660 stronger for electron-electron compared to electron-proton interactions. The key issue for how important such a process would be is related to whether the interaction distance is sufficiently small at the current densities at which OMR is observed.

The triplet-polaron interaction (TPI) model for OMR [7] suggests that, like many other excitonic models, the primary effect of the magnetic field is to change the balance between singlets and triplets. This may happen through an interaction at the exciton level, but an exchange at the pair state level would give the same result [7-9]. The net effect of this interaction is to produce more singlets (thus improving efficiency) whilst reducing the triplet population. The TPI model suggests that it is this reduction in the triplet population that is responsible for the relative change in current (i.e. the OMR is a secondary effect). The initial assumption for this mechanism is that changes in the triplet population change the mobility of the polarons, which has been experimentally verified by dark injection measurements in poly-(3-hexylthiophene) (P3HT) layers with either Au/P3HT/Au or Au/P3HT/Al structures [15]. In that work, it was suggested that if a polaron had the same spin state as the corresponding state on the triplet exciton then the site would be effectively blocked to transport and hence mobility would be decreased. In addition, if the polaron had an opposite spin state to the corresponding state on the triplet then the polaron could interact with that triplet and there would either be a scattering event, resulting in a triplet and polaron, or a quenching event which would leave only the polaron. Again, either of these processes would have some interaction time and would be expected to result in a decrease in mobility.

This model of OMR suggests that the interaction of polarons with triplet excitons is important, which implies that the polarons are either adjacent to or on the molecule in the triplet state. The current density at which we first see evidence of OMR in our devices, $\sim 0.01$ A/cm$^2$, corresponds to $\sim 6 \times 10^{12}$ electrons/s.cm$^2$. Given that Alq$_3$ has an areal density of $\sim 1.6 \times 10^{14}$ molecules/cm$^2$ we can calculate the ratio of the areal density of polarons per second to Alq$_3$ molecules to be $\sim 0.01$ electrons per second per molecule. Given that for a typical device thickness the active layer is $\sim 100$ molecules thick this means that even at the lowest current density at which we observe OMR there is a high probability of a polaron visiting any molecular site. Since typical *operating* current densities are of the order of 1-10 A/cm$^2$, interactions between polarons and any molecular site are inevitable. We therefore suggest that as hyperfine coupling to proton spins has been ruled out, interactions between paramagnetic species (polarons) and excited states (or pair states) may be responsible for decreasing the triplet concentration and hence increase the efficiency and current density in the device, although we cannot rule out other more exotic scenarios [e.g 22].

**Acknowledgements**

We thank the EPSRC National Mass Spectrometry Service Centre, Swansea for the analysis of Alq$_3$-$d_{18}$. This work was conducted with financial support from the Leverhulme Trust (AJD).

**References**

1). J. Kalinowski, M. Cocchi, D. Virgili, P. Di Marco, and V. Fattori, Chem. Phys. Lett. 380, 710, (2003).

2). T. L. Francis, O. Mermer, G. Veeraraghavan, and M. Wohlgenannt, New Journal of Physics 6, 185 (2004).

3). Ö. Mermer, G. Veeraraghavan, T. L. Francis, and M. Wohlgenannt, Solid State Comm. 134, 631, (2005).

4). Ö. Mermer, G. Veeraraghavan, T. L. Francis, Y. Sheng, D. T. Nguyen, M. Wohlgenannt, A. Köhler, M. K. Al-Suti, and M. S. Khan, Phys. Rev. B 72, 205202, (2005).

5). V.N. Prigodin, J.D. Bergeson, D.M. Lincoln, and A.J. Epstein, Synthetic Metals 156, 757-761 (2006).

6). Yue Wu, Zhihua Xu, Bin Hu, and Jane Howe, Phys. Rev. B 75, 035214-6 (2007).

7). P. Desai, P. Shakya, T. Kreouzis, W. P. Gillin, N. A. Morley, and M. R. J. Gibbs, Phys. Rev. B 75, 094423-5 (2007).

8). P. Desai, P. Shakya, T. Kreouzis, and W. P. Gillin, J. Appl. Phys. 102, 073710-5 (2007).

9). P. Desai, P. Shakya, T. Kreouzis, and W. P. Gillin, Phys. Rev. B 76, 235202-5 (2007).

10). P. A. Bobbert, T. D. Nguyen, F. W. A. van Oost, B. Koopmans, and M. Wohlgenannt, Phys. Rev. Lett. 99, 216801-4 (2007).

11). Y. Sheng, T. D. Nguyen, G. Veeraraghavan, O. Mermer, M. Wohlgenannt, S. Qiu, and U. Scherf, Phys. Rev. B 74, 045213-9 (2006).

12).    P. A. Bobbert, W. Wagemans, F. W. A. van Oost, B. Koopmans, and M. Wohlgenannt, Phys. Rev. Lett. 102, 156604-4 (2009).

13).    T.D. Nguyen, Y. Sheng, M. Wohlgenannt, and T.D. Anthopoulos, Synthetic Metals 157, 930-934 (2007).

14).    P. Shakya, P. Desai, T. Kreouzis, and W. P. Gillin, J. Appl. Phys. 103, 043706-5 (2008).

15).    J. Y. Song, N. Stingelin, W. P. Gillin, and T. Kreouzis, Appl. Phys. Lett., 93, 233306 (2008).

16).    C.C. Tong and K.C. Hwang, J. Phys. Chem. C 111, 3490-3494 (2007).

17).    P. Shakya, P. Desai, M. Somerton, G. Gannaway, T. Kreouzis, and W. P. Gillin, J. Appl. Phys. 103, 103715-5 (2008).

18).    J. M. Brown and A. Carrington (2003). *Rotational spectroscopy of diatomic molecules*. Cambridge: Cambridge University Press. 137-138.

19). A.J. Drew, J. Hoppler, L. Schulz, F.L. Pratt, P. Desai, P. Shakya, T. Kreouzis, W.P. Gillin, A. Suter, N.A. Morley, V.K. Malik, A. Dubroka, K.W. Kim, H. Bouyanfif, F. Bourqui, C. Bernhard, R. Scheuermann, G.J. Nieuwenhuys, T. Prokscha and E. Morenzoni, Nature Materials 8, 109 - 114 (2009).

20) W. G. Richards, H. P. Trivedi and D. L. Cooper, Spin-orbit coupling in molecules, Clarendon Press, Oxford, 1981.

21) R. Batra, J. Phys. Chem. 100, 18371-18379 (1996).

22) D. Hsieh, Y. Xia, L. Wray, D. Qian, A. Pal, J. H. Dil, J. Osterwalder, F. Meier, G. Bihlmayer, C. L. Kane, Y. S. Hor, R. J. Cava, and M. Z. Hasan, Science 323, 919-922 (2007)

Figure captions

Figure 1.         (a) The current density in the 90 nm Alq$_3$ (circles) and Alq$_3$-$d_{18}$ (triangles) devices as a function of drive voltage. The inset to the figure shows the photoluminescence spectra of the two materials recorded under identical conditions.

Figure 2.         The percentage change in current density (OMR) as a function of magnetic flux density for a 90 nm Alq$_3$ (circles) and Alq$_3$-$d_{18}$ (triangles) devices at different drive voltages.

Figure 3.         The percentage change in efficiency as a function of magnetic flux density for a 90 nm Alq$_3$ (circles) and Alq$_3$-$d_{18}$ (triangles) devices at different drive voltages.

Figure 1

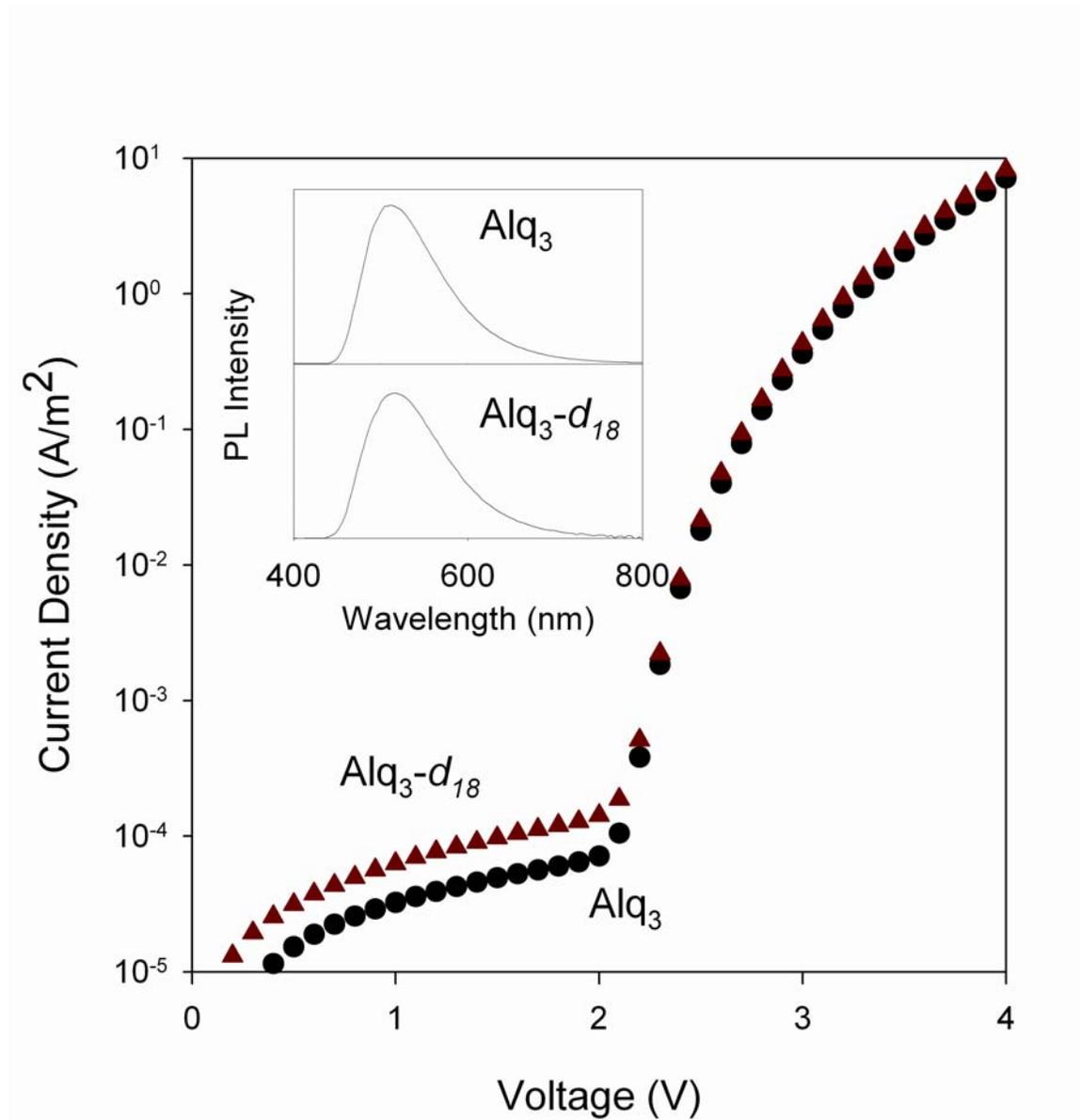

Figure 2

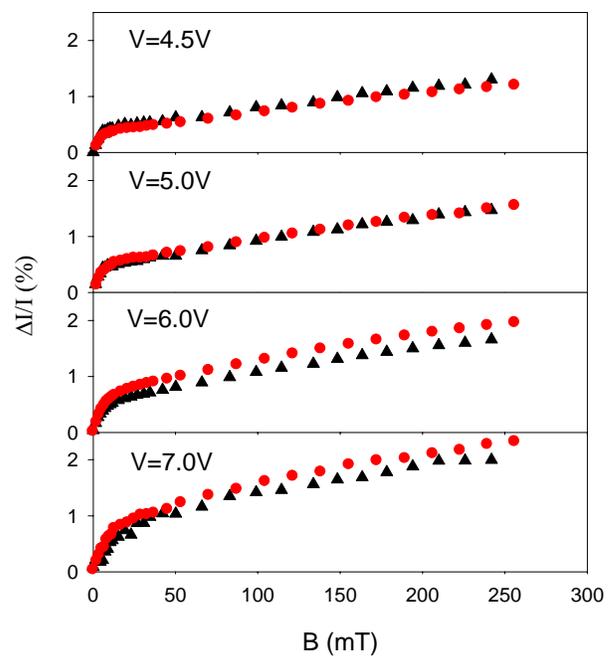

Figure 3

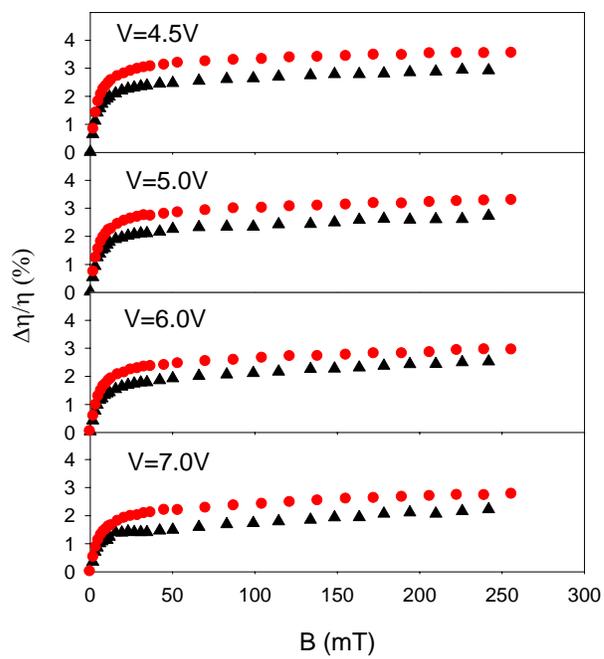